\documentclass[dvipdf twoside, letterpaper, 11pt]{article}
\RequirePackage[round]{natbib}
\setcitestyle{aysep={,}}
\usepackage{graphicx,psfrag}
\usepackage{epsfig}
\usepackage[colorlinks=true,linkcolor=blue,citecolor=blue,urlcolor=blue]{hyperref}
\usepackage{amsmath, amsthm, amssymb, bbm, subfig, mathrsfs, enumerate}
\usepackage{amsmath,amsfonts,bbold}
\usepackage{setspace}
\usepackage{float}
 \usepackage{relsize}
\def\E{\mathbb E}
\def\P{\mathbb P}
\def\V{\mathrm{Var}}
\def\es{{\scriptstyle[\emptyset]}}

\newcommand{\comment}[1]{}
\defcitealias{asaguidelines}{ASA, 2014}

\begin{document}

\bibliographystyle{plainnat}

\title{Teaching Statistics at Google Scale\footnote{This article is derived from a talk presented at the Joint Statistical Meetings on August 5, 2013, in Montreal, during the session entitled {\it Toward Big Data in Teaching Statistics}.}}
\author {
  Nicholas Chamandy\thanks{N. C. is a Data Scientist at Lyft.} \and
  Omkar Muralidharan\thanks{O. M. is a Statistician in Ads Quality at Google.} \and
  Stefan Wager\thanks{S. W. is a Ph.D. Candidate in Statistics at Stanford University.}
}

\maketitle

\begin{abstract}
Modern data and applications pose very different challenges from those of the 1950s or even the 1980s.
Students contemplating a career in statistics or data science need to have the tools
to tackle problems involving massive, heavy-tailed data, often interacting with live, complex systems.
However, despite the deepening connections between engineering and modern data science, we argue that
training in classical statistical concepts plays a central role
in preparing students to solve Google-scale problems. To this end, we present
three industrial applications where significant modern data challenges were overcome by statistical thinking.
\end{abstract}

\section{Introduction}

Technology companies like Google generate and consume data on a
staggering scale. Massive, distributed data present novel and
interesting challenges for the statistician, and have spurred much
excitement among students, and even a new discipline: Data
Science. Hal Varian famously quipped in 2009 that Statistician would
be ``the sexy job in the next 10 years'' \citep{nyt}, a claim
seemingly backed up by the proliferation of job postings for data
scientists in high-tech. The McKinsey Global Institute took a more urgent tone
in their 2011 report examining the explosion of big data in industry
\citep{mckinsey}. While extolling the huge productivity gains that
untapping such data would bring, they predicted a shortage of hundreds of thousand of ``people with
deep analytical skills'', and millions of data-savvy managers, over the
next few years. 

Massive data present great opportunities for a
statistician. Estimating tiny experimental effect sizes becomes
routine, and practical significance is often more elusive than mere
statistical significance. Moreover, the power of approaches that pool
data across observations can be fully realized. But these exciting opportunities come with strings attached. The data sources and structures that a graduating statistician (or data scientist) faces are unlike those of the 1950s, or even the 1980s. Statistical models we take for granted are sometimes out of reach: constructing a matrix of dimensions $n$ by $p$ can be pure fantasy, and outliers are the rule, not the exception. Moreover, powerful computing tools have become a prerequisite to even reading the data. Modern data are in general unwieldy and raw, often contaminated by `spammy' or machine-generated observations. More than ever, data checking and sanitization are the domain of the statistician.

In this context, some might think that the future of data science
education lies in engineering departments, with a focus on building
ever more sophisticated data-analysis systems. Indeed, the American
Statistical Association Guidelines Workgroup noted the ``increased
importance of data science'' as the leading Key Point of its 2014
Curriculum Guidelines \citepalias{asaguidelines}. As Diane Lambert and
others have commented, it is vital that today's statisticians have the
ability to ``think with data'' \citep{hardin}. We agree wholeheartedly
with the notion that students must be fluent in modern computational
paradigms and data manipulation techniques. We present a counter-balance to
this narrative, however, in the form of three data analysis challenges
inspired by an industrial `big data' problem: click cost estimation. We
illustrate how each example can be tackled not with fancy computation,
but with new twists on standard statistical methods, yielding solutions that are not only
principled, but also practical.

Our message is not at odds with the ASA's recent recommendations;
indeed, the guidelines highlight statistical theory, ``flexible problem
solving skills'', and ``problems with a substantive context'' as core
components of the curriculum \citepalias{asaguidelines}. The methodological tweaks presented in this article are not particularly advanced, touching on well-known results in the domains of resampling, shrinkage, randomization and causal inference. Their contributions are more conceptual than theoretical. As such, we believe that each example and solution would be accessible to an undergraduate student in a statistics or data science program. In relaying this message to such a student, it is not the specific examples or solutions presented here that should be stressed. Rather, we wish to emphasize the value of a solid understanding of classical statistical ideas as they apply to modern problems in preparing tomorrow's students for large-scale data science challenges.

In many cases, the modern statistician acts as an interface between the raw data and the consumer of those data. The term `consumer' is used rather broadly here. Traditionally it would include the key decision-makers of a business, and perhaps the scientific research community at large. Increasingly, computerized `production' systems---automated decision engines that are critical to a company's success---may also be counted among the consumers. Leaning on this framing, we see that difficulties may arise in all three phases of data science: the input phase, the inference phase, and the output phase.

Section \ref{sec:mr} highlights some challenges posed by the input or data retrieval phase. We describe a common paradigm for data reduction and some of the constraints it imposes, and then show how we can overcome the resulting engineering difficulties with a targeted use of the bootstrap.
In Section \ref{sec:tail} we present a concrete modeling problem made more difficult by the scale of the data, and discuss how to judiciously simplify the problem to alleviate the computational burden without needlessly reducing the quality of inference.
Section \ref{sec:fb} considers the output phase---specifically, the situation where the output of a statistical model is piped directly into a production system. We frame this in terms of a causal inference paradigm, highlight a problem which may result, and propose a remedy. In addition to providing evidence for the value of statistical analysis in modern data science, we also hope that these examples can provide students with an idea of the kind of problems that modern industrial statisticians may be confronted with.

\section{Sharding, MapReduce, and the Data Cube}
\label{sec:mr}
Many new challenges in statistics arise from the size and structure of modern data sources.
Here we focus on difficulties caused by a sharded architecture, where pieces of the data are stored on
many different machines (or shards), often in a manner determined by system efficiency rather than ease
of analysis. Such architectures usually arise when we need to work with data sources that are too large to be
practically handled on a single machine. In practice, the sharding happens far upstream of any data analysis;
the statistician typically has no control over which pieces of data are stored on which machines.

\begin{figure}[t]
  \centering
  \includegraphics[width=1\textwidth]{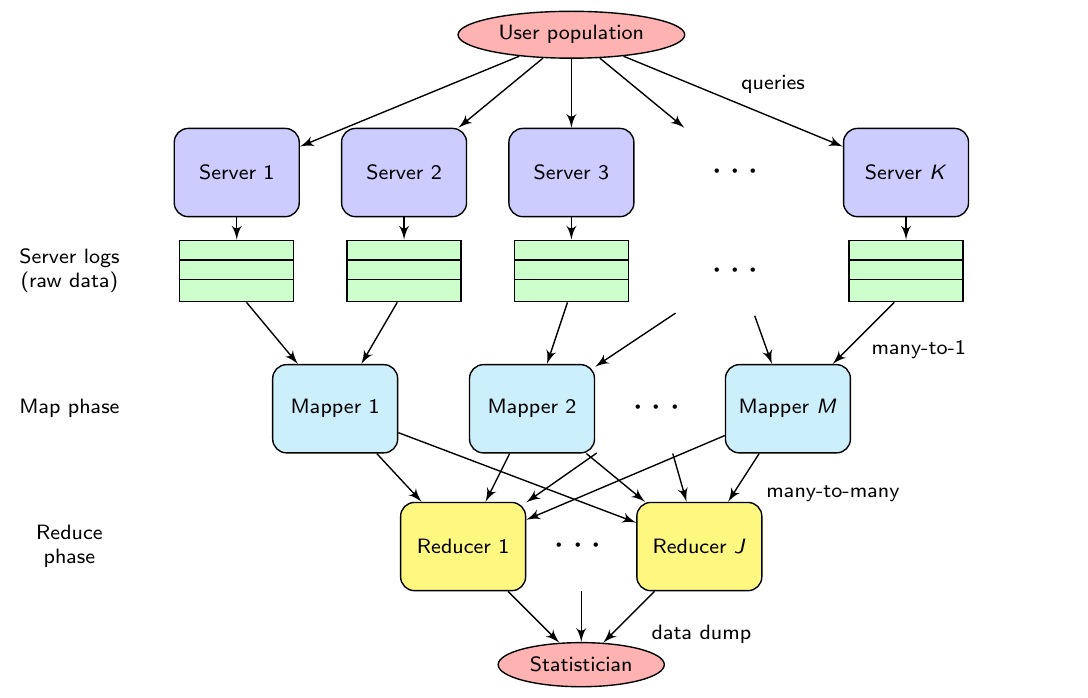}
 \caption{A stylized example of a Google data stream. Every
    user-generated query is processed by one of $K$ servers, with
    different queries from the same user possibly hitting different
    servers. The servers write raw data to logs, one record per query.
    These records are first processed by $M$ mapper machines, whose output is
    shuffled along the key of interest $k$ and mapped in order to $J$
    reducer machines. The data
    are further processed by the reducers into a single summary tuple
    for each unique value of $k$. These data are finally analyzed by
    the statistician. Typically, $K\gg M \gg J$.}
  \label{fig:cartoon}
\end{figure}

One common procedure for reducing massive data to a smaller form
amenable to inference is the MapReduce framework \citep{mr}.
A detailed description of MapReduce is beyond the scope of this paper;
however we give a brief overview in Section \ref{sec:mr_review} below.

At a high level, MapReduce lets us create (multi-way) histograms of our data.
This computational architecture makes computing some functionals of the
data easy, while other functionals are more difficult to get. For example, if $x \in \mathbb{R}^{100}$, then computing
$\mathbb{E}[x]$ is easy, but computing the number of pairs of observations that
are exact duplicates of each other is hard. As statisticians, we need to express the primitives on
which we base our inferential procedures in terms of queries that are feasible with MapReduce.

\subsection{Computing with Map Reduce}
\label{sec:mr_review}

Here, we follow the exposition of \cite{stream}, who consider a basic instance of MapReduce;
reproduction of a stylized diagram of the system is provided in Figure
\ref{fig:cartoon}. The internals of MapReduce are less important to this
discussion than the sort of data that the program takes as input and
spits out at the other end as output. The input can in general be petabytes
($10^{15}$B) of sharded, unstructured or hierarchical data
objects (such as machine-generated log files). The output is typically a
flattened table of key-value tuples of aggregated statistics.

Most MapReduce jobs run in just a few hours (or less). In its simplest form, the
MapReduce network consists of a single {\it master} machine, which coordinates
the process, and many {\it mapper} and {\it reducer} machines. The system
is fully parallel in that there is no communication among mappers nor
among reducers (though there is some limited communication between the two
groups, and both communicate with the master). 

During the {\it  map phase}, each mapper processes its set of input records, which may
themselves be complex non-vector objects, and produces an intermediate
tuple by applying some aggregation function. The tuples are different from
the final output because they have been computed from only a subset of
the data, and must be further combined. These intermediate data are then sorted by their key and
sent as input to the reducers, a process known as
shuffling. In the {\it reduce phase}, each reducer machine further aggregates its own input
tuples to produce a final statistic for every unique key. The resulting data, which are of a manageable size, can then be downloaded from the MapReduce network and interrogated using R, SQL, and other such tools. Their final structure is what is known as a data cube: a set of key-value pairs where the key is a possibly high-cardinality categorical vector, and the value is a highly aggregated numeric vector. For example, in a Google Search application, the key might be the cartesian product of all countries, languages, web browser versions, types of computing device, and hours of day. The value might contain aggregated statistics like the number of queries and clicks in each bin.

\subsection{Example: Variance estimation with big data}
\label{sec:var}
In the context of MapReduce, uncertainty estimation can be challenging, even for relatively rudimentary statistics. The reason is that the record used for sharding and storing the data rarely coincides with the unit of analysis, i.e., the smallest unit for which an independence assumption can be made. Therefore, computing exact second-order statistics like sums of squares is impossible without expensive communication between machines in the network.


Consider the following example. We work for a pay-per-click advertising platform (such as Google),
and wish to estimate the average cost per ad click paid by advertisers over some collection of search queries in different countries.
We get to measure the cost of \emph{clicks}, each of which can be
attributed to a \emph{user} who is in one of the \emph{countries} of interest.
Specifically, let $Y_{iuc}$ denote the cost of the $i$-th click of the $u$-th user in
the country $c$. Finally, suppose that the ``record units'' are clicks, meaning that each click is stored separately,
and there are no guarantees about whether any pair of clicks may or may not be on the same shard. All clicks are not, in general, independent.

Although the data may be massive, obtaining a simple point estimator for the mean country-wise cost per click
is easily accomplished using MapReduce, with sums as the aggregation functions, clicks and spend as the values,
and country as the key.
Writing $N_c$ for the total number of clicks in country $c$, we can estimate our parameter
of interest as
\begin{equation}
  \hat\theta_c = Y_{\cdot \cdot  c} / N_c, \text{ where } Y_{\cdot \cdot c} = \sum_{i, \, u}Y_{iuc}.
\end{equation}
As statisticians, however, we are rarely satisfied with a simple point estimate $\hat\theta$, and also
want a measure of the stability of this estimator. And this is where the details of the MapReduce implementation
become important.

Given that the data are stored at the click-level, we can easily compute
a na\"{i}ve variance estimate using MapReduce,
\begin{equation}
\label{eq:click_var}
  S_c^2 = \frac{1}{N_c}\left(\frac{1}{N_c} \sum_{i, \, u} Y_{ijc}^2 - \hat\theta_c^2\right),
\end{equation}
by simply tacking on a sum of squared click costs, $Y_{iuc}^2$, to the value tuple. More explicitly, when mapping over a given click, the MapReduce program first determines its country, say $c$, followed by its cost $Y_{iuc}$, and then emits the tuple $(1, Y_{iuc}, Y_{iuc}^2)$ to the entry of a summation aggregator with country key $c$.
The variance estimator $S_c^2$ would be justified under the assumption that all click records in the data are independent.

In general, however, the estimator $S_c^2$ is biased, because the true ``experimental unit'' is something coarser than
the click. In particular, $S_c^2$ will fail to account for the (usually positive) correlation between
the cost of successive clicks from the same user. In a typical computing architecture,
we can easily get around this issue by computing user-level variances:
\begin{equation}
\label{eq:user_var}
V_c^2 = \frac{1}{M_c}\left\{\left(\frac{M_c}{N_c}\right)^2\frac{1}{M_c} \left(\sum_{u} Z_{uc}^2\right) - \hat\theta_c^2\right\}; \text{ where }
Z_{uc} = \sum_{i} Y_{iuc}
\end{equation}
and $M_c$ is the number of unique users in country $c$. The problem is, the individual $Y_{iuc}$ comprising
user $u$'s total cost $Z_{uc}$ are not usually on the same shard, and so we need to compute a separate MapReduce query to get each $Z_{uc}$ (or key our aggregators by $u$). Thus, to compute \eqref{eq:user_var}, the size of the output returned by MapReduce would need to scale as the number of users in the system. As the number of users is almost as large as the number of clicks, this is computationally prohibitive.

\subsection{The Poisson bootstrap}

Perhaps surprisingly, however, it turns out that we can side-step the \emph{engineering} problem of
how to compute $V_c^2$ with a \emph{statistical} idea. Although $V_c^2$ itself is difficult to compute,
we can produce a MapReduce-friendly alternative to it using a variation of the bootstrap \citep{efron1994introduction}. The key trick is to replace multinomial bootstrap sampling with Poisson sampling, which can be
implemented while streaming over the data.

More specifically, recall that bootstrap procedures draw a sequence of resamples which can be viewed
as weighted versions of the original sample. The idea behind the Poisson bootstrap is to use Poisson
weight vectors instead of multinomial weight vectors, since the former can be generated on-the-fly
without knowing the sample size, while the latter cannot. The method works as follows.
To construct $B$ replicates of the output tuple from a MapReduce, $B$ independent {\sf Poisson}(1)
random variables are drawn for each record in the data, as the record is processed. Given a record
belonging to a statistical unit with the unique identifier $u$, the random number generator for replicate $b$ is seeded with
$g(h(u) + b)$, where $g$ and $h$ are two fingerprint functions\footnote{A fingerprint function $f$,
as used here, is a mapping of arbitrary data into the space of 64-bit integers. It is virtually 1-to-1,
so that collisions are extremely rare (but not impossible). Moreover, knowing $f(x_1), f(x_2) ,\ldots, f(x_n)$,
for some arbitrary collection of points $x_1, \, \ldots, \, x_n$, gives no information about the relationship
between those data points.}. This procedure ensures that any two records originating from the same
unit will have identical Poisson vectors. The $b$-th Poisson variable represents a count of how many
times the unit is to appear in replicate $b$.

If there are $n$ total units, the sample size of any Poisson bootstrap replicate is not fixed, but rather
a random variable distributed as {\sf Poisson}$(n)$. Nevertheless, the Poisson bootstrap is
asymptotically equivalent to the traditional multinomial bootstrap, and to streaming analogues of related resampling techniques \citep{stream}. These ideas have been studied by, among others, \citet{han-mac}, \citet{lee-clyde}, and \citet{politis1999subsampling}.

\subsection{Illustration with Google data}

\begin{figure}[t]
\centering
  \includegraphics[trim=0cm 1.5cm 0cm 0cm, clip=true, width=0.85\textwidth]{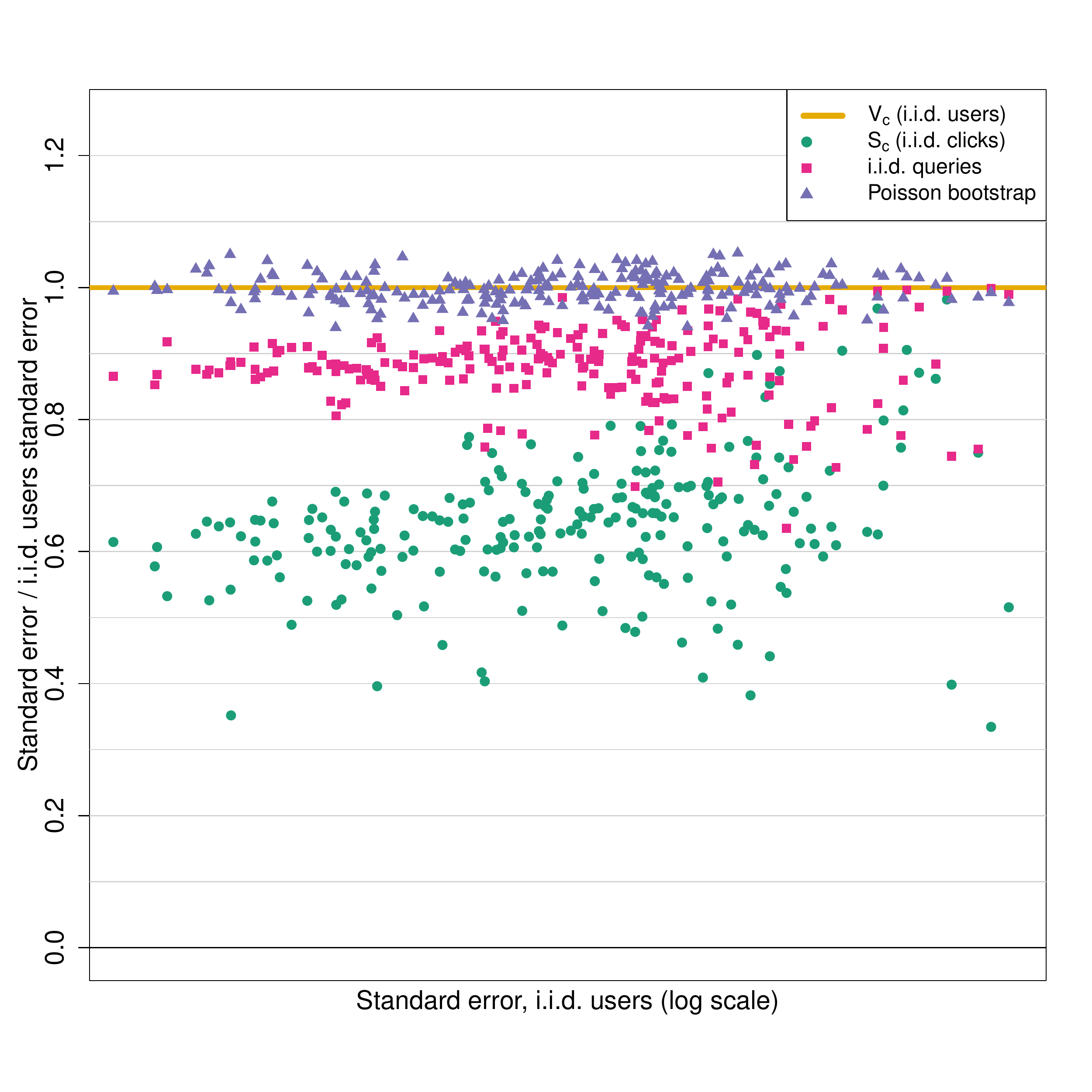}
 \caption{Estimated standard deviation of average click cost in 234 countries. Standard errors are computed relative to the computationally prohibitive independent-user estimator $V_c$ (gold line and x-axis, on a log scale). Assuming that all clicks are i.i.d. leads to $S_c$, which underestimates uncertainty  (green dots). So does, to a lesser extent, assuming that all queries are i.i.d. (pink squares). The gold standard $V_c$ is fairly well-approximated in this example by applying the Poisson bootstrap with 1000 replicates (purple triangles).}
\comment{
Half confidence intervals for mean click cost in 32 randomly chosen countries, extending from 0 to 1.96 $\times$ the estimated standard error. Assuming that all clicks are independent (green) leads to an underestimation of uncertainty, as does, to a lesser extent, assuming that all queries are independent (pink). The computationally prohibitive independent-user calculation (gold) is a more accurate reflection of the true random process. It is well-approximated in this example by applying the Poisson bootstrap with 1000 replicates (purple).}
  \label{fig:ci}
\end{figure}

Figure \ref{fig:ci} illustrates this problem with Google data, by showing standard errors for average
click cost in 234 countries. The data are normalized so that estimates are displayed relative to $V_c$ \eqref{eq:user_var}, which is based on an i.i.d. user assumption.\footnote{Computing $V_c$ for this problem required using a very expensive data join, which would not be feasible as a part of a routine data-processing pipeline.} As we might have feared, the na\"ive variance estimates $S_c^2$ \eqref{eq:click_var}, which assume that clicks are i.i.d., are badly biased downwards. We also include an intermediate estimator computed from an i.i.d. query model; in general, a query should be thought of as a set of actions comprising a single task that may include many clicks. Such a model is tractable in certain applications; query-level sharding, as depicted in Figure \ref{fig:cartoon}, is not uncommon. Accounting for the within-query correlation helps, but the i.i.d. query estimator still underestimates uncertainty on the standard error scale by about 10\%.

Poisson bootstrap standard errors using $B=1000$ replicates
are also displayed in Figure \ref{fig:ci}. They are quite close, at least in expectation,
to the gold standard $V_c$ computed from a ``random user'' model---that is, using between-user sums of squares. This is as expected, since the bootstrap procedure assumes that users are the independent units of analysis. Thus, the Poisson bootstrap has successfully provided us with a computationally efficient way of
estimating the variance of $\hat{\theta}_c$ using MapReduce.

\section{The Long Tail}
\label{sec:tail}

Large data sets are notorious for having long-tailed occurrence distributions:
many units occur infrequently, but together, these infrequent units
account for a large fraction of total events. For example, Google
advertisers have collectively input millions of keywords into their
advertising campaigns. These keywords are a targeting mechanism---they
allow the advertiser to indicate a desire to show an ad when the
user's search query contains the specified terms. But most Google
ad keywords are only seen a small number of times. Because of this, when doing statistical analysis of such long-tailed data,
the number of units of interest scales with the amount of data collected.
This creates computational challenges for traditional statistical
methods.

\subsection{Example: Cost per click modeling}

Suppose we are interested in predicting the average cost per ad click
on each keyword. Such a prediction model might serve a number of
purposes: classifying query strings into commerciality bins,
estimating the revenue that will be generated by a new ad campaign,
etc. Some of these uses may entail memory or computational constraints, for
example if the model will be used as part of an online serving system.

As a simple version of this, we might want to fit a linear mixed effects model
of the form
\[
\log\left(\mbox{ad click cost}\right)\sim\mbox{country}+\mbox{query category}+\left(1|\mbox{keyword}\right),
\]
where `query category' refers to some coarse classification of query
strings into topics like {\it Home and Garden}, {\it Finance}, {\it
  Arts and Entertainment}, and so on. Here we have expressed the model
using an R {\tt lme4}-style model formula \citep{bates2014lme4}. The
response variable is ad click cost on the log scale. The last term on
the right-hand-side denotes a simple keyword random effect, i.e., a
different mean 0 random offset for every keyword. The fixed effects of
country and query category are assumed to be constant across all
keywords. This may be overly simplistic if a query category encompasses a diverse set of products, but is a reasonable first
attempt. More sophisticated random effects models are of course
possible but don't fundamentally change the problem.

For convenience, we can fit the model in two stages: first, fit a
coarse model without the keyword effect; and second, analyze the residuals
to estimate keyword-level random effects. Fitting the coarse model
is easy---a straightforward regression on categorical predictors with
which any undergraduate statistician should be familiar. The
sufficient statistics are also particularly simple: total clicks and log cost
for each combination of country and query category. These statistics
are not only easy to
compute using MapReduce, but also easy to work with afterward, fitting
nicely in memory (as long as there aren't too many query categories). 

Next, we need to analyze the keyword-level residuals. This is where
the challenge lies. There are many keywords, most of which are extremely
rare. Intuitively, though, the rare keywords are not completely
useless. \comment{there is information to be
extracted from any multiply-occurring keyword. What's more, w}Ideally,
we would
like to be able to `borrow strength' from the distribution of keyword
effects when estimating parameters for any given keyword. This is the
key idea behind shrinkage estimators. In a second MapReduce step, one can easily apply the first stage model
and even collect keyword-level residual statistics; for simplicity, a count, mean
and sample variance. 

Ignoring the uncertainty in the
variances, the random effects problem then reduces to that of
estimating a vector of normal means. Suppose each keyword $k$ has a true effect $\mu_{k}$. We observe a noisy
sample mean residual $R_{k}$, with true variance $\sigma^2_k$ inversely proportional to its frequency:
\begin{eqnarray*}
\mu_{k} & \sim & \textrm{N}\left(0,\eta^{2}\right)\\
\sigma_{k} & \sim & G\\
R_{k}|\mu_{k},\sigma_{k} & \sim & \textrm{N}\left(\mu_{k},\sigma_{k}^{2}\right),
\end{eqnarray*}
where $G$ is an unspecified prior on the keyword-level standard
deviations. We want to estimate $\mu_{k}$ based on $R_{k}$ and
$\sigma_{k}$.

We could choose to treat all keywords equally. But for most applications, it is more important to perform well on frequent
keywords. Therefore, our loss function must be weighted by frequency
(or inverse variance). This is no different from treating every
unique instance of a keyword as a separate observation. The
appropriate squared-error loss is then 
\begin{equation}\label{eq:loss}
\mathcal{L}\left(\mu,\hat{\mu}\right)=\sum_k \mathcal{L}_k\left(\mu,\hat{\mu}\right) = \sum_k\left(\mu_{k}-\hat{\mu}_{k}\right)^{2}/\sigma_{k}^{2}.
\end{equation}

We wish to minimize risk, that is the expected loss
$\E[\mathcal{L}\left(\mu,\hat{\mu}\right)]$ averaged over the
priors. Assuming known $\eta$ and $\sigma^2_k$'s, this is achieved by using the Bayes
estimator \citep{lehmann1998}, i.e. the posterior mean
\begin{eqnarray*}
\mathbb{E}\left(\mu_{k}|R_{k},\sigma_{k}\right) & = & \frac{\eta^{2}}{\sigma_{k}^{2}+\eta^{2}}R_{k}.
\end{eqnarray*}
This estimator shrinks the observed average residuals towards 0 (the prior
mean). For the $k$-th keyword, the Bayes estimator has risk 
\begin{align*}
\mathcal{R}_{\sf Bayes} &= \E\left(\mu_{k} - \E\left(\mu_{k}|R_{k},\sigma_{k}\right)\right)^2/\sigma_k^2\\
&= \E\left[\V(\mu_k | R_{k},\sigma_{k})/\sigma_k^2\right]\\
&= \E\left[\eta^2/(\sigma_k^2 + \eta^2)\right]\\
&= \E\left[\tau_k/\left(1+\tau_k\right)\right],
\end{align*}
where $\tau_k=\eta^{2}/\sigma^{2}_k$ is the signal to noise ratio, and
the last expectation is taken over $G$. This estimator minimizes our loss
function, but it is computationally expensive. Not only does it require computing
statistics for all keywords, but more importantly storing
a coefficient for each keyword if we wish to make predictions from the
model. The latter would be especially problematic in an online
setting, where speed and memory are at a premium. Nevertheless, this `long tail' problem gives a hint as to how
we might proceed. In practice, we often wish to make predictions for examples we
have never before seen, corresponding to $\tau_k=0$ and
therefore $\hat\mu_k=0$ (there is no signal, only noise). Intuitively,
making a prediction for cases we have only observed once or twice, of
which there may be millions, is a qualitatively similar task.

\subsection{Pruning rare units}

The Bayes-optimal approach described above spends its computational resources unwisely---it uses the same amount of memory (one coefficient) for each keyword.
This is wasteful since the rare keywords are less important in the
loss function, and their effects are harder to estimate
anyway. Nevertheless, the Bayes
estimator is a clear improvement on our prior knowledge. Compared to simply plugging $\hat\mu \equiv \E(\mu_k)=0$ into \eqref{eq:loss}, it reduces the $k$-th term in our risk by
\begin{align*}
\E[\mathcal{L}_k(\mu, 0)] - \E\left[\mathcal{L}_k(\mu, \E(\mu_k |
  R_k. \sigma_k))\right] & =
                           \E\left[\frac{\mbox{Var}\left(\mu_{k}\right)-\mbox{Var}\left(\mu_{k}|R_{k},\sigma_{k}\right)}{\sigma_{k}^{2}}\right] \\
& = \E\left[\frac{\eta^{2}}{\sigma_{k}^{2}}\left(\frac{\eta^{2}}{\eta^{2}+\sigma_{k}^{2}}\right)\right]\\
& = \E\left[\tau_k\left(\frac{\tau_k}{1+\tau_k}\right)\right].
\end{align*}
Recall that $\tau_k$ is proportional to the frequency of keyword
$k$. For rare keywords, $\tau_k$ is small, so the Bayes estimator is
only a small improvement over the prior in such cases.

This suggests an obvious strategy: compute the Bayes estimate for
common keywords, and use the prior for rare keywords. This can be
implemented by running a MapReduce to find keywords for which $\tau_{k}$
is bigger than some threshold $T$, gathering data for these keywords
(either in the same MapReduce or in another one), and using the truncated
estimator
\begin{equation}\label{eq:mu-trunc}
\hat{\mu}_{k}=\begin{cases}
\frac{\eta^{2}}{\sigma_{k}^{2}+\eta^{2}}R_{k} & \tau_{k}\geq T\\
0 & \tau_{k}<T
\end{cases}.
\end{equation}

A similar calculation shows that for keyword $k$, the risk of this estimator is
\[
\mathcal{R}_{\sf Trunc}=\mathbb{E}\left(\frac{\tau_k}{1+\tau_k}\left(1+\tau_k\mathbbm{1}\left\{ \tau_k<T\right\} \right)\right).
\]

The truncated estimator is often nearly as good as the Bayes
estimator. Clearly, $\mathcal{R}_{\sf Trunc}\leq\left(1+T\right)\mathcal{R}_{\sf Bayes}$. If most keywords are rare
and only a few occur frequently, $\tau$ is concentrated near 0, making
$\tau_k\mathbbm{1}\left\{ \tau_k<T\right\}$ small and $\mathcal{R}_{\sf
  Trunc}$ close to $\mathcal{R}_{\sf Bayes}$. Also, since even a small
$T$ is usually enough to eliminate a huge number of keywords, the
truncated estimator can perform nearly as well as the Bayes estimator
while using a small fraction of the keywords. A nice side-benefit in online applications is that
\eqref{eq:mu-trunc} provides us with an initial click cost estimate
even for keywords that we have never before observed.

Extensions of similar pruning ideas to more general contexts such as
regression and optimization have been considered by, among others,
\citet{duchi2010composite}, \citet{langford2009sparse}, and
\citet{mcmahan2013ad}. Of course, in a real application $\eta$ and
$G$, and even perhaps the prior mean, would be
unknown. While a classical Bayesian might hazard an educated guess at
these, perhaps bolstered by diagnostic checks, in the big data regime
one can do better. {\it Empirical Bayes} provides a framework for
estimating these parameters from the data \citep{casella, efron2010large}.
While these ideas are especially powerful in a big data setting, they are
surprisingly underrepresented in typical undergraduate curricula.

\subsection{Simulation study}

\begin{figure}[t]
  \centering
  \includegraphics[trim=0cm 0cm 0cm 0cm, clip=true, width=1\textwidth]{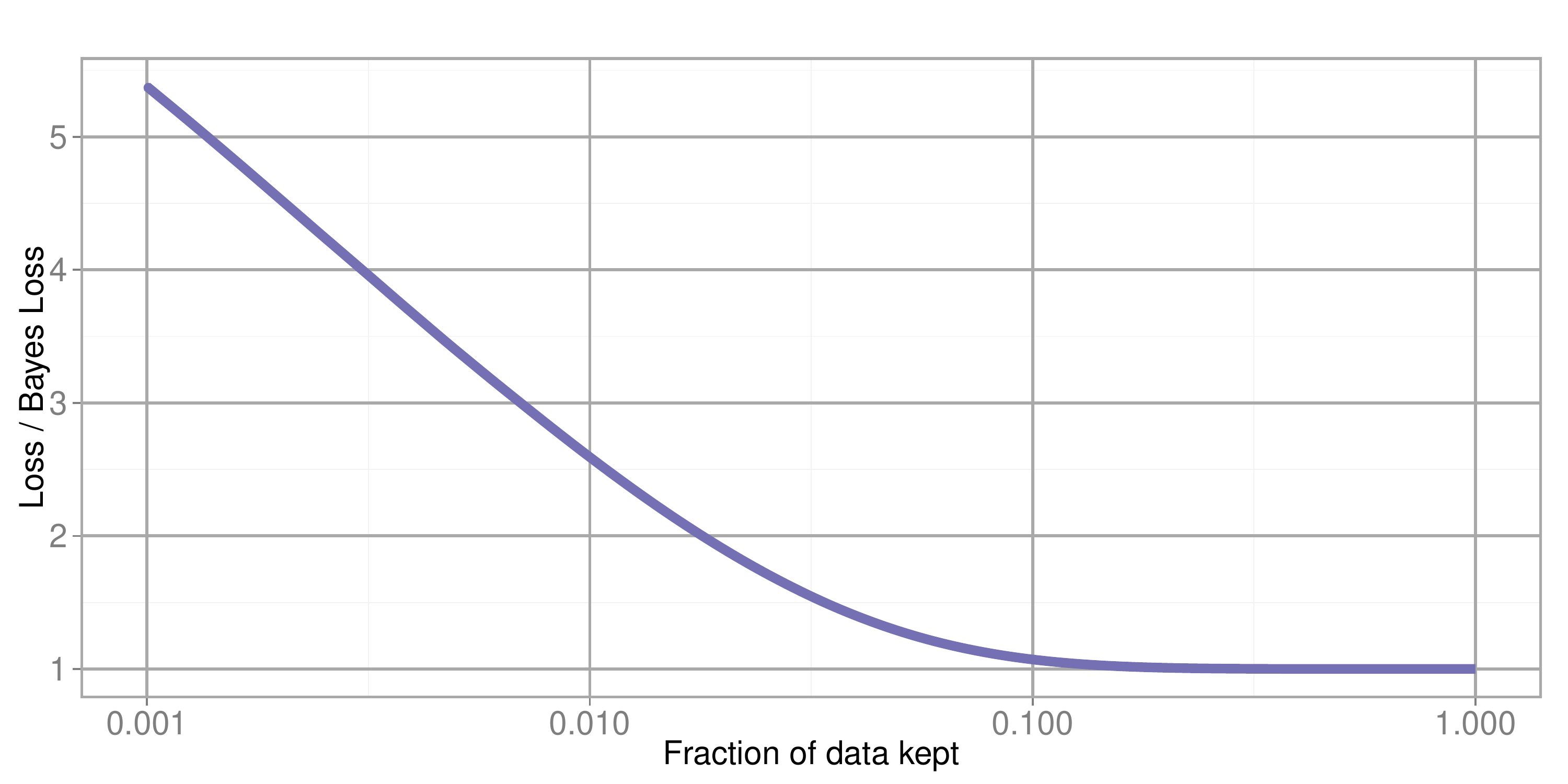}
 \caption{Simulated loss incurred by the truncated estimator in the rare keyword example, relative to Bayes loss, for different occurrence thresholds $T$. The x-axis indicates the fraction of all keyword effects estimated (i.e., $\P[\tau \ge T]$). By estimating only 10\% of the random effects, and discarding the rest of the keywords, we perform about 7.5\% worse than optimal while obtaining an order-of-magnitude computational savings. }
 \label{fig:omkar} 
\end{figure}

Figure \ref{fig:omkar} illustrates with a simulation study how the truncated estimator described in the previous section can offer attractive performance-computation tradeoffs. We generated
$\sigma_{k}^{2}$ from a log-normal distribution $G$ with $\mbox{sd}\left(\log\left(\sigma_{k}\right)\right)=1.5$,
and set $\eta=\mathbb{E}\left(1/\sigma_{k}^{2}\right)=1$. The far
right shows that if we keep all the data, the truncated and Bayes
estimators are the same, so $\mathcal{R}_{\sf Trunc}/\mathcal{R}_{\sf Bayes}=1$.
As we increase the threshold $T$, we start filtering out keywords,
and move left on the plot. $\mathcal{R}_{\sf Trunc}/\mathcal{R}_{\sf Bayes}$
rises, but slowly---we can keep just $10\%$ of the data and perform
only $7.5\%$ worse than the Bayes estimator. Finally, as we start
discarding informative keywords, $\mathcal{R}_{\sf Trunc}/\mathcal{R}_{\sf Bayes}$
explodes and we perform much worse than the Bayes estimator.

\section{Statistical Feedback}
 \label{sec:fb}
 
In traditional applications, both academic and industrial, statistical models or inference procedures tend to be of a `throw-away' nature. A model is fit and a parameter estimated, or a decision is made under uncertainty, and the result is communicated to the relevant stakeholders. Typically, the statistician's job ends there, and the model or inference is unlikely to be revisited except if improvements to the methodology are desired, or in rare follow-up or meta analyses.

In this section we describe a scenario
where the output of a sequentially-run statistical procedure is at risk of being used by one or more of its consumers in a way that could be damaging to the inference itself. Such use is typically not malicious, nor can it really be termed abuse; nonetheless it may have unintended side-effects.

\subsection{Example: Query classification in ads serving}

Suppose that we have built a statistical classifier to predict whether a search query has commercial intent.
Our goal is to use the predictions from our ``commerciality'' classifier for stratified data analysis, forecasting query growth, or some other such offline application.
These predictions are updated by periodically collecting feature data over some time window, but not by retraining the model itself (or by retraining it only very rarely);
the reason for this is that fitting our commerciality model is costly as it requires collecting hand-labeled training data and curation by a statistician.

In this example, we chose to use expected click cost, i.e., the output from the model in Section \ref{sec:tail}, as a feature in our classifier. If we were operating in isolation, this choice would be completely innocuous.
However, if we share the output of our classifier with other teams, feedback loops can emerge. For example, suppose that another team---the ads serving engineering team---hears of the existence of our model for predicting commercial intent, and wants to exploit that signal in their own algorithms.
Perhaps, for example, they wish to search for more candidate ads on queries that they deem to be highly commercial. If our model is logistic regression and produces probabilities $\hat Y_i = \hat\P[$query $i$ is commercial$]$, this may correspond to situations where $\hat Y_i>p$, for some threshold $p$. Alternatively they may wish to fetch a set of candidates whose size is proportional to the odds of $\hat Y$.
 
On the surface, such uses of our model are entirely reasonable. The problem is, if the other engineering team uses our signal in a way that changes the expected click cost, then our model may get biased by feedback. For example, take a query for which our model predicts a large probability, either correctly or because of some prediction error. The serving system may suddenly begin to fetch more candidate ads for this query, increasing auction pressure and leading to higher average click costs. Since we haven't retrained our classifier in the interim, this increase in click cost will be taken by the model as evidence that the query is even more likely to be commercial than previously thought. This effect can worsen in successive time periods. On the other hand, a query for which we erroneously predict a small $\hat Y$ may be disfavoured in serving, pushing down click cost and reinforcing our misguided belief that the query lacks commercial intent. 

\subsection{Detecting feedback with noise injection}

Although feedback itself is an engineering problem, detecting feedback is a statistics problem.
Perhaps the most difficult part of getting a handle on statistical feedback is to define precisely and
quantitatively what we mean by feedback. Here, we follow the approach of \citet{nips}, and frame
the problem in terms of the Rubin Causal Model \citep{rcm} and counterfactual model predictions.

More specifically, let  $\hat Y_i^t\es$ denote the predictions we would have made for the
$i$-th item (query, say) in time period $t$ if there hadn't been any feedback, i.e., the prediction
that our model would make if the engineering team never began using it in production.
Other counterfactual predictions are indexed by the prediction deployed in the previous period:
 $\hat Y^{t+1}_i{\scriptstyle[\hat Y_i^t]}$ denotes the time $t+1$ model prediction,
in a system with feedback, given that $\hat Y_i^t$ was the prediction for item $i$ produced at time $t$.
Given this notation, we define the feedback at time $t+1$ on item $i$ as a difference in potential outcomes,
\begin{equation}
\label{eq:feedback}
\text{feedback}_i^t = \hat Y^{t+1}_i{\scriptstyle[\hat Y_i^t]} - \hat Y_i^{t+1}\es,
\end{equation}
i.e., the difference between the prediction the system in fact made, and the prediction it would have made were there no feedback.

\newcommand\independent{\protect\mathpalette{\protect\independenT}{\perp}}
\def\independenT#1#2{\mathrel{\rlap{$#1#2$}\mkern2mu{#1#2}}}

In order to use the formalism \eqref{eq:feedback} in practice, we need
to restrict the kinds of feedback we seek to detect. Our first
assumption is that feedback is \emph{Markovian}, i.e., that
$\text{feedback}_i^t$ is conditionally independent of $\hat Y_i^{1},
\, ..., \, \hat Y_i^{t - 1}$ given $\hat Y_i^t$.
Second, we assume that the feedback is \emph{additive}, i.e., that it
does not depend on $\hat Y_i^{t+1}\es$ conditional on $\hat Y_i^t$. 

Given these two assumptions, we can define a \emph{feedback function}
\begin{align}\label{eq:ff}
f(y) &= \mathbb{E}\left[\hat Y^{t+1}_i{\scriptstyle[\hat Y_i^t]} - \hat Y_i^{t+1}\es  \, \big| \, \hat Y_i^t = y \right] \\
&= \mathbb{E}\left[\hat Y^{t+1}_i{\scriptstyle[\hat Y_i^t]} - \hat Y_i^{t+1}\es \, \big| \, \hat Y_i^t = y, \, \hat Y_i^{t+1}\es, \, \hat Y_i^{1}, \, ..., \, \hat Y_i^{t - 1} \right].
\end{align}
The goal of our analysis is to estimate this feedback function $f(\cdot)$.
The principal challenge in doing so is that it is not possible
to simultaneously observe the two sets of predictions represented by the two terms on the right-hand side of
\eqref{eq:ff}, namely $\hat Y^{t+1}_i{\scriptstyle[\hat Y_i^t]}$ and $\hat Y_i^{t+1}\es$.

The solution we proposed in \cite{nips} involves secretly injecting a noise term $\nu_i^t$ into the prediction
at time $t$. The intuition behind this approach is that it creates a sort of synthetic randomized experiment,
allowing us to estimate the average causal effect $f(y)$ of feedback. Noise injection is related to the idea of instrumental
variables \citep{ang-imb-rub} in the sense that, once we have injected this artificial noise into the system,
we analyze its effect on the system just like we would in an instrumental variables regression.

For simplicity, consider the case where feedback enters into the model linearly
\begin{equation}\label{eq:lin-f}
\hat Y^{t+1}_i{\scriptstyle[\hat Y_i^t]} = \hat Y_i^{t+1}\es + \theta \, \hat Y_i^t, \text{ and so } f(y) = \theta \, y.
\end{equation}
This model is rather unlikely to hold in reality, but it admits a simple analysis. Note that the formulation \eqref{eq:lin-f} implicitly assumes that all the randomness in the system has been absorbed into $\hat Y^{t+1}_i\es$.

Now, in order to detect the feedback, suppose that we start secretely deploying noisy predictions $\hat Y_i^t+\nu_i^t$ instead of $\hat Y_i^t$ in serving at time $t$. In other words, we never tell other engineering teams what the raw value $\hat Y_i^t$ produced by our predictor is, and only give them access to the noised-up quantity $\hat Y_i^t+\nu_i^t$. Any feedback mechanism must now act through these noisy predictions, and so the relationship \eqref{eq:lin-f} becomes
\begin{align}
  \hat Y^{t+1}_i{\scriptstyle[\hat Y_i^t+\nu_i^t]} &= \hat Y_i^{t+1}\es + f(\hat Y_i^t+\nu_i^t) \\
&= \hat Y_i^{t+1}\es + \theta \hat Y_i^t + \theta\nu_i^t. \label{eq:reg}
\end{align}
Here, both the new model predictions $\hat Y^{t+1}_i{\scriptstyle[\hat Y_i^t+\nu_i^t]}$ and the noise terms $\nu_i^t$ are observed. Thus, since $\nu_i^t$ is independent of all other quantities on the right-hand side of \eqref{eq:reg} by construction, we can estimate $\theta$ from \eqref{eq:reg} using simple linear regression of $\hat Y^{t+1}{\scriptstyle[\hat Y^t+\nu^t]}$ on $\nu^t$.

Under this recipe, the precision with which we estimate the feedback slope is a ratio of the variance of the noise we have injected to the variance of the noise inherent in our prediction model:
\begin{equation}\label{eq:th-var}
  n\V(\hat\theta) = {\V\left(\hat Y_i^{t+1}{\scriptstyle[\hat Y_i^t]}\right)} \, \big/ \, {\V(\nu_i^t)}.
\end{equation}
There is a clear tradeoff. Adding more noise improves our ability to
measure feedback, but of course it degrades the quality of the
published predictions that will be used in production. More details
pertaining to \eqref{eq:th-var}, and extensions of this idea to general non-linear feedback functions, are considered in \citet{nips}.

Having an estimate of $\theta$ or, more generally, $f(\cdot)$, can be useful in several ways.
First, merely detecting non-zero feedback can alert us to emerging problems and help us identify ``rogue'' uses of our published signals $\hat Y^i_t$.
More ambitiously, we could also try to do a post hoc feedback correction. For example, in the case of linear feedback, we could use an estimate $\hat\theta$ to generate sanitized predictions $\hat Y^{t+1}_i\stackrel{\textrm{def}}{=}\hat Y^{t+1}_i{\scriptstyle[\hat Y_i^t+\nu_i^t]} - \hat\theta(\hat Y_i^t+\nu_i^t)$; our simulation experiments below use this correction.

\subsection{Simulation study}

To illustrate this technique, we simulate from a simplistic setting: linear feedback in a linear model. This allows us to demonstrate concretely how the randomization procedure might be carried out in ideal circumstances, and even to approximately remove the offending feedback once we've detected it. 

We start with the following true underlying model:
\begin{equation}\label{eq:lin}
  Y_i^t = \alpha + \beta X_i^t + \epsilon_i^t, \;\;\;  \epsilon_i^t \sim \textrm{N}(0, \sigma^2).
\end{equation}
We assume that enough data is collected when we train the model at $t=0$ so that $\alpha$ and $\beta$ can be taken as known (since the interesting source of error is in the feedback, not in the estimation itself). We imagine that we must make $n$ weekly predictions for a full year before our next opportunity to retrain the model; i.e., $T=52$. The predictors $X$ are generated as follows:
\begin{equation}\label{eq:x-cf}
  X_i^{t+1}{\scriptstyle [\hat Y_i^t]} = X_i^{t+1}\es + \gamma \hat Y_i^{t},
\end{equation}
where
\begin{equation}
  X_i^{t+1}\es = X_i^{t}\es + \delta Y_i^{t} + \xi_i^t, \;\;\; \xi_i^t \sim \textrm{N}(0, \tau^2).
\end{equation}
Thus the predictors vary across time even before the feedback occurs, in a way that may depend on the true value of $Y$. Combining \eqref{eq:lin} and \eqref{eq:x-cf}, it can be seen that the feedback function is $f(y) = \theta y$ with $\theta = \gamma\beta$.

\begin{figure}[t]
  \centering
  \includegraphics[trim=0cm 0cm 0cm 0cm, clip=true, width=0.7\textwidth]{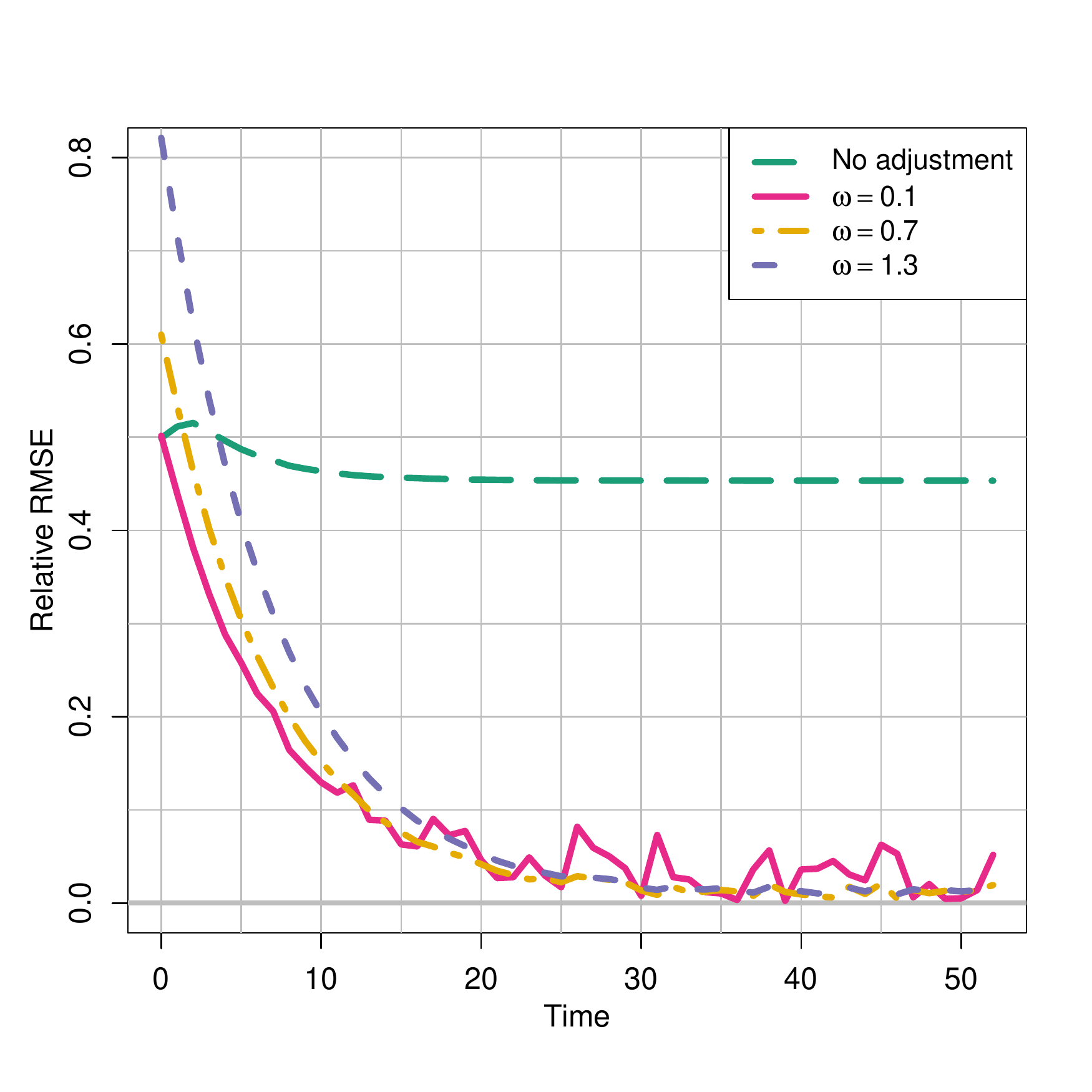}
 \caption{Simulated relative root-mean square error across 52 time points for linear feedback in a 1-predictor linear model. If no adjustment to our predictor is made, feedback causes an asymptotically (in time) constant amount of error in our predictions (green dashed line). The remaining lines show the error obtained by injecting normal noise of different standard deviations $\omega$, and then using it to remove an estimate of the feedback from $X$ before making a new prediction.}
 \label{fig:fb-lin} 
\end{figure}

Given a set of simulated predictions $\hat Y_i^t$ at time $t$, we first add noise $\nu_i^t \sim \textrm{N}(0, \omega^2)$. We then generate a new set of $X$s from \eqref{eq:x-cf}, but with $\hat Y_i^{t} + \nu_i^t$ in place of $\hat Y_i^{t}$. Next we\comment{compute the differences
\begin{equation}
  Z_i^{t+1} = X_i^{t+1}{\scriptstyle [\hat Y_i^t]} - X_i^{t}\es = \delta Y_i^t + \gamma\hat Y_i^t +\gamma\nu_i^t + \xi_i^t,
\end{equation}
and then} perform the regression of $X^{t+1}{\scriptstyle [\hat Y^{t}+\nu^{t}]}$ on $\nu^t$. \comment{
\begin{equation}
  Z^{t+1} \sim 1 + \hat Y^t + \nu^t.
\end{equation}}
The coefficient of $\nu^t$ from this regression is our estimate of $\gamma$, call it $\hat\gamma^{t+1}$. We then compute the corrected prediction
\begin{equation}
  \hat Y_i^{t+1} = \beta \left(X_i^{t+1}{\scriptstyle [\hat Y_i^t + \nu_i^t]} - \hat\gamma^{t+1} (\hat Y_i^{t}+\nu_i^t)\right),
\end{equation}
again `deploy' the noised-up version
$\hat Y_i^{t+1} + \nu_i^{t+1}$, 
and iterate until $t=T$.

Note that in this example the variables and predictions grow in expectation across time. Therefore, in Figure \ref{fig:fb-lin} we plot the {\it relative} root-mean squared error of our estimator, defined as
\begin{equation} 
  R(t) = \left(\E[(\hat Y_i^t - Y_i^t)^2]/\E[Y_i^t]^2\right)^{1/2} ,
\end{equation}
as a function of time. We allowed the standard deviation of the injected noise, $\omega$, to vary, while fixing the remaining parameter values: $\beta=1$, $\sigma=1$, $\gamma=0.3$, $\delta=0.15$, $\tau=1$, $n=10^5$.

The general effectiveness of the procedure, along with the tradeoff inherent in the choice of $\omega$, is evident from the curves in Figure \ref{fig:fb-lin}.
With $\omega=0.1$, early predictions have about as much error as na\"ive predictions; this error decays quickly but rather noisily. Noise with $\omega=0.7$ seems about right for this problem. When we inject too much noise, e.g. $\omega=1.3$, we begin with poor predictions and don't really make up for it with more stability at larger $t$.

\section{Conclusions}

In this paper, we described three sub-problems within the broader context of click cost estimation for paid web search. By no means have we presented an exhaustive list of methodological challenges awaiting the modern industrial statistician. Rather, we have chosen the problems to highlight a few distinct ways in which modern data and applications differ from traditional statistical applications, yet can benefit from careful statistical analysis.

In the first problem, uncertainty estimation was nontrivial because the types of second-order statistics that statisticians normally take for granted can be prohibitively expensive in large data streams. The second problem, which centered around building a fine-grained prediction model, explored the gains that are possible by judiciously chopping off the long tail of a massive data set. Finally, Example 3 highlighted the importance of protecting one's methods from unexpected uses, in an environment where product decisions driven by statistical models can feed back inconspicuously into the inference procedure. 
All three problems initially appeared to be straightforward and amenable to standard methods, until a technical wrinkle presented itself. In each case, it was primarily through careful statistical reasoning, and {\em not} by designing some heavyweight engineering machinery, that we were able to arrive at a solution.

In doing so, we drew upon ideas that have been around for at least
four decades---and in some cases much longer. Over that time, these
methods have been taught to undergraduate students with an eye to
small data problems. It is increasingly important, in light of the
rapid growth of available data sources, to modernize classroom
examples so that graduating students are not shell-shocked by their
first encounter with big data. Likewise, data-based simulation techniques should be emphasized
as a proving-ground for methodology before applying it na\"{i}vely to a billion data points. Early exposure to tools like MapReduce
can help build confidence, but frankly, a newly-minted statistician with ``deep
analytical skills'' should have no trouble learning such things on the
job. As we have argued, it remains as critical as ever that we continue to equip students with classical techniques, and that we teach each and every one of them to think like a statistician.


\subsection*{Acknowledgements}
The authors are grateful to Amir Najmi, Hal Varian, and three
anonymous reviewers for their many helpful suggestions. We would also
like to thank Nick Horton, Jo Hardin and Tim Hesterberg for
encouraging us to write the manuscript.

\bibliography{tas}

\begin{thebibliography}{20}
\providecommand{\natexlab}[1]{#1}
\providecommand{\url}[1]{\texttt{#1}}
\expandafter\ifx\csname urlstyle\endcsname\relax
  \providecommand{\doi}[1]{doi: #1}\else
  \providecommand{\doi}{doi: \begingroup \urlstyle{rm}\Url}\fi

\bibitem[Angrist et~al.(1996)Angrist, Imbens, and Rubin]{ang-imb-rub}
J.~D. Angrist, G.~W. Imbens, and D.~B. Rubin.
\newblock Identification of causal effects using instrumental variables.
\newblock \emph{Journal of the American Statistical Association}, 91\penalty0
  (434):\penalty0 444--455, 1996.

\bibitem[{Bates} et~al.(2014){Bates}, {M{\"a}chler}, {Bolker}, and
  {Walker}]{bates2014lme4}
D.~{Bates}, M.~{M{\"a}chler}, B.~{Bolker}, and S.~{Walker}.
\newblock {Fitting Linear Mixed-Effects Models using lme4}.
\newblock \emph{ArXiv e-prints}, June 2014.

\bibitem[Casella(1985)]{casella}
G.~Casella.
\newblock An introduction to {E}mpirical {B}ayes data analysis.
\newblock \emph{The American Statistician}, 39\penalty0 (2):\penalty0 83--87,
  1985.

\bibitem[Chamandy et~al.(2012)Chamandy, Muralidharan, Najmi, and Naidu]{stream}
N.~Chamandy, O.~Muralidharan, A.~Najmi, and S.~Naidu.
\newblock Estimating uncertainty for massive data streams.
\newblock \emph{Internal technical report, Google}, 2012.
\newblock
  [\href{https://static.googleusercontent.com/media/research.google.com/en/us/pubs/archive/43157.pdf}{link}].

\bibitem[Dean and Ghemawat(2004)]{mr}
J.~Dean and S.~Ghemawat.
\newblock {M}ap{R}educe: Simplified data processing on large clusters.
\newblock In \emph{Proceedings of the 6th Symposium on Operating Systems Design
  \& Implementation}, pages 137--149, 2004.

\bibitem[Duchi et~al.(2010)Duchi, Shalev-Shwartz, Singer, and
  Tewari]{duchi2010composite}
J.~Duchi, S.~Shalev-Shwartz, Y.~Singer, and A.~Tewari.
\newblock Composite objective mirror descent.
\newblock In \emph{Conference on Learning Theory}, 2010.

\bibitem[Efron(2010)]{efron2010large}
B.~Efron.
\newblock \emph{Large-Scale Inference: Empirical {B}ayes Methods for
  Estimation, Testing, and Prediction}.
\newblock Cambridge University Press, 2010.

\bibitem[Efron and Tibshirani(1993)]{efron1994introduction}
B.~Efron and R.~Tibshirani.
\newblock \emph{An Introduction to the Bootstrap}.
\newblock CRC press, 1993.

\bibitem[Hanley and MacGibbon(2006)]{han-mac}
J.~A. Hanley and B.~MacGibbon.
\newblock Creating non-parametric bootstrap samples using {P}oisson
  frequencies.
\newblock \emph{Computer Methods and Programs in Biomedicine}, 83:\penalty0
  57--62, 2006.

\bibitem[{Hardin} et~al.(2014){Hardin}, {Hoerl}, {Horton}, and {Nolan}]{hardin}
J.~{Hardin}, R.~{Hoerl}, N.~J. {Horton}, and D.~{Nolan}.
\newblock {Data Science in the Statistics Curricula: Preparing Students to
  ``Think with Data''}.
\newblock \emph{ArXiv e-prints}, October 2014.

\bibitem[Langford et~al.(2009)Langford, Li, and Zhang]{langford2009sparse}
J.~Langford, L.~Li, and T.~Zhang.
\newblock Sparse online learning via truncated gradient.
\newblock \emph{Journal of Machine Learning Research}, 10:\penalty0 719--743,
  2009.

\bibitem[Lee and Clyde(2004)]{lee-clyde}
H.~K.~H. Lee and M.~A. Clyde.
\newblock Online {B}ayesian bagging.
\newblock \emph{The Journal of Machine Learning Research}, 5:\penalty0
  143--151, 2004.

\bibitem[Lehmann and Casella(1998)]{lehmann1998}
E.~Lehmann and G.~Casella.
\newblock \emph{{Theory of Point Estimation}}.
\newblock Springer Verlag, 1998.
\newblock ISBN 0387985026.

\bibitem[Lohr(2009)]{nyt}
S.~Lohr.
\newblock For today's graduate, just one word: Statistics.
\newblock \emph{The New York Times}, 2009.
\newblock
  [\href{http://www.nytimes.com/2009/08/06/technology/06stats.html?\_r=0}{link}].

\bibitem[Manyika et~al.(2011)Manyika, Institute, Chui, Brown, Bughin, Dobbs,
  Roxburgh, and Byers]{mckinsey}
J.~Manyika, McKinsey~Global Institute, M.~Chui, B.~Brown, J.~Bughin, R.~Dobbs,
  C.~Roxburgh, and A.H. Byers.
\newblock \emph{Big Data: The Next Frontier for Innovation, Competition, and
  Productivity}.
\newblock McKinsey Global Institute, 2011.
\newblock URL \url{https://books.google.com/books?id=vN1CYAAACAAJ}.

\bibitem[McMahan et~al.(2013)McMahan, Holt, Sculley, Young, Ebner, Grady, Nie,
  Phillips, Davydov, Golovin, et~al.]{mcmahan2013ad}
H.~B. McMahan, G.~Holt, D.~Sculley, M.~Young, D.~Ebner, J.~Grady, L.~Nie,
  T.~Phillips, E.~Davydov, D.~Golovin, et~al.
\newblock Ad click prediction: A view from the trenches.
\newblock In \emph{Proceedings of the 19th ACM SIGKDD international conference
  on Knowledge discovery and data mining}, pages 1222--1230. ACM, 2013.

\bibitem[Politis et~al.(1999)Politis, Romano, and Wolf]{politis1999subsampling}
D.~N. Politis, J.~P. Romano, and M.~Wolf.
\newblock \emph{Subsampling}.
\newblock Springer Series in Statistics. Springer New York, 1999.

\bibitem[Rubin(2005)]{rcm}
D.~B. Rubin.
\newblock Causal inference using potential outcomes.
\newblock \emph{Journal of the American Statistical Association}, 100\penalty0
  (469):\penalty0 322--331, 2005.

\bibitem[Wager et~al.(2014)Wager, Chamandy, Muralidharan, and Najmi]{nips}
S.~Wager, N.~Chamandy, O.~Muralidharan, and A.~Najmi.
\newblock Feedback detection for live predictors.
\newblock In \emph{Advances in Neural Information Processing Systems}, 2014.

\bibitem[Workgroup(2014)]{asaguidelines}
American Statistical Association Undergraduate~Guidelines Workgroup.
\newblock 2014 {\it curriculum guidelines for undergraduate programs in
  statistical science}.
\newblock Alexandria, VA: American Statistical Association, 2014.
\newblock
  [\href{http://www.amstat.org/education/curriculumguidelines.cfm}{link}].

\end{thebibliography}

\end{document}